\documentclass[twocolumn,amsmath,amssymb,color,superscriptaddress,footinbib,longbibliography]{revtex4-2}
\usepackage{stix}
\usepackage[colorlinks,citecolor=blue,linkcolor=blue,urlcolor=blue]{hyperref}
\usepackage{graphicx}
\usepackage{dcolumn}
\usepackage{multirow}
\usepackage{bm}
\usepackage{footmisc}
\usepackage{xcolor}
\usepackage{float}
\usepackage[english]{babel}
\usepackage{url}
\usepackage{ulem}
\usepackage{titlesec}
\usepackage{tabularx}
\usepackage{times}
\usepackage{cases}

\begin{document}
\title{Bridging distributed quantum materials via multi-hotspot vacuum: remote Cooper pairing and Andreev teleportation}

\author{Zuzhang Lin}
\affiliation{New Cornerstone Science Lab, Department of Physics, The University of Hong Kong, Hong Kong, China}
\affiliation{HK Institute of Quantum Science \& Technology, The University of Hong Kong, Hong Kong, China}
\affiliation{State Key Laboratory of Optical Quantum Materials, The University of Hong Kong, Hong Kong, China}

\author{Zhijian Song}
\affiliation{New Cornerstone Science Lab, Department of Physics, The University of Hong Kong, Hong Kong, China}
\affiliation{HK Institute of Quantum Science \& Technology, The University of Hong Kong, Hong Kong, China}
\affiliation{State Key Laboratory of Optical Quantum Materials, The University of Hong Kong, Hong Kong, China}

\author{Chengxin Xiao}
\affiliation{New Cornerstone Science Lab, Department of Physics, The University of Hong Kong, Hong Kong, China}
\affiliation{HK Institute of Quantum Science \& Technology, The University of Hong Kong, Hong Kong, China}
\affiliation{School of Electrical and Information Engineering, Zhengzhou University, Zhengzhou, Henan 450001, China}

\author{Wang Yao}
\email{wangyao@hku.hk}
\affiliation{New Cornerstone Science Lab, Department of Physics, The University of Hong Kong, Hong Kong, China}
\affiliation{HK Institute of Quantum Science \& Technology, The University of Hong Kong, Hong Kong, China}
\affiliation{State Key Laboratory of Optical Quantum Materials, The University of Hong Kong, Hong Kong, China}

\date{\today}

\begin{abstract}
We introduce an architecture where mesoscopic quantum matter distributed over spatially separated nodes can be correlated in equilibrium, creating an unprecedented form of many-body quantum system.
Central to the design is a multi-gap split-ring resonator (SRR) where the cavity photon has multiple hot spots---each with deep-subwavelength volume at a split gap and separated by millimeter-scale distances. The cavity's vacuum fluctuations can then mediate a many-body interaction that bridges the distributed quantum materials embedded in the multiple gaps, coupling them into a single correlated mesoscopic system in equilibrium.
As an example, we consider a THz SRR with two split gaps, each proximitized to a metallic moir\'e superlattice, where virtual exchange of a photon in the cavity vacuum mediates a current--current interaction across the gaps. The inherent attractive interaction channels lead to remote Cooper pairing reminiscent of mesoscopic superconductivity, demonstrated with density matrix renormalization group and exact diagonalization calculations.
With the two constituents of a Cooper pair now paired across a millimeter-scale separation, a hole incident at one split gap can be converted into an outgoing electron at the remote gap, a process we term Andreev teleportation. The entanglement entropy between the two mesoscopic superlattices is shown to scale linearly with the area (total number of sites) of the mesoscopic lattice.
Our results suggest an intriguing paradigm for equilibrium quantum networks of mesoscopic matter that enable emergent nonlocal functionalities and distributed quantum resources.
\end{abstract}
\maketitle
Many-body interactions are the engine of emergence in condensed matter. They correlate and entangle particles, and from these arise collective phenomena---superconductivity, magnetism, topological order---with no analogue in non-interacting systems.
Realizing new forms of interaction that are not available in natural materials can open the door to previously inaccessible realms of correlated physics.
A key paradigm in modern quantum physics is using confined photons---for instance, in microcavities and waveguides---to mediate many-body interplay between atoms or artificial atoms (qubits), enabling the engineering of novel quantum matter \cite{mivehvar_cavity_2021,chang_colloquium_2018,blais_circuit_2021,schlawin_cavity_2022,ritsch_cold_2013}.
To circumvent the limitation in light--matter coupling strength, a prevailing approach exploits Raman-type processes where atoms scatter photons between applied laser fields and the cavity mode~\cite{pellizzari_decoherence_1995,imamoglu_quantum_1999,dimer_proposed_2007,baumann_dicke_2010,gao2020photoinduced}.
An effective nonlocal interaction between any pair of atoms within the cavity volume can then be established through the exchange of a virtual cavity photon, with coupling strength controlled by the laser~\cite{imamoglu_quantum_1999,dimer_proposed_2007,baumann_dicke_2010,gao2020photoinduced}.
In an alternative operating regime, two such scattering processes can be combined via the exchange of a real photon propagating from one cavity to another ~\cite{cirac_quantum_1997,yao_theory_2005}
to entangle two remote qubits as a resource for quantum network~\cite{duan_long-distance_2001,yao_theory_2005}.


Recent advances in cavity quantum electrodynamics have enabled a new regime of quantum light--matter interaction~\cite{forn-diaz_ultrastrong_2019,frisk_kockum_ultrastrong_2019,bloch_strongly_2022}, where the ultrastrong coupling strength comparable to transition frequencies, can efficiently renormalize light--matter ground states through virtual excitations. This has opened a new frontier for manipulating matter using the vacuum field of an empty cavity without any external illumination~\cite{garcia-vidal_manipulating_2021,schlawin_cavity_2022,lu_cavity_2025,paravicini-bagliani_magneto-transport_2019, keller_few-electron_2017, scalari_ultrastrong_2012, chen_review_2016, paravicini-bagliani_gate_2017, jeannin_ultrastrong_2019}.
Experiments have demonstrated a plethora of vacuum-modified phenomena such as magneto-transport~\cite{paravicini-bagliani_magneto-transport_2019}, integer~\cite{appugliese_breakdown_2022} and fractional quantum Hall effects~\cite{enkner_tunable_2025}, chemical reactions~\cite{thomas_tilting_2019}, metal--insulator transitions~\cite{jarc_cavity-mediated_2023}, and superconductivity~\cite{thomas_exploring_2025}. 
In parallel, theoretical work has extended cavity vacuum control to an even broader range of phenomena---among them charge density wave~\cite{li2020manipulating}, ferroelectricity~\cite{ashida_quantum_2020}, 
topological phase transitions~\cite{lin_remote_2023,wei_cavity-vacuum-induced_2025}, symmetry breaking~\cite{jiang_quantum_2019,hubener_engineering_2021,ke_vacuum-induced_2023,lin_spontaneous_2026},
electron--phonon couplings~\cite{sentef_cavity_2018} and superconductivity~\cite{li2020manipulating,lu_cavity-enhanced_2024,kozin_cavity-enhanced_2025, curtis_cavity_2019}. 
Notably, virtual photon exchange between 2D electrons has been proposed as a pairing mechanism for superconductivity mediated solely by the cavity vacuum field~\cite{schlawin2019cavity}, although the superconducting instability at thermodynamic limit has been debated~\cite{andolina_amperean_2024}.


\begin{figure*}
\centering
\includegraphics[width=0.9\textwidth]{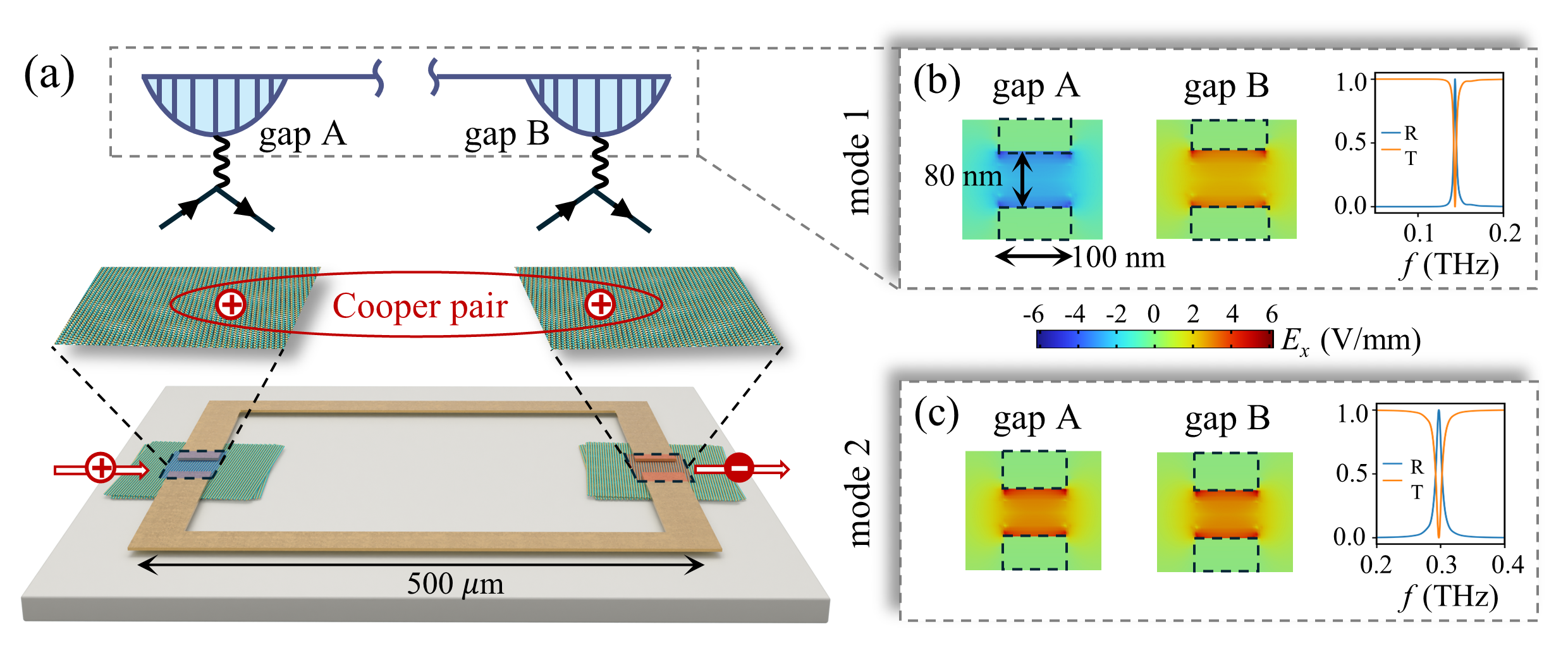}
\caption{ \textbf{Distributed mesoscopic matter at multiple hotspots of a cavity mode, bridged via its vacuum fluctuations}.
\textbf{(a)} Schematic of a metallic ring THz resonator with two split gaps, each proximitized by a moir\'{e} superlattice. A cavity mode has a spatial profile spread across both gaps, and its vacuum fluctuations can mediate an attractive interaction between carriers across the two gaps, leading to remote Cooper pairing.
This pairing enables an {\it Andreev teleportation} process: injection of a hole into gap A results in an electron ejection from gap B.
\textbf{(b), (c)} Spatial profiles and resonance spectra of the two lowest cavity modes for an exemplary square ring resonator where the two split gaps are separated by $L=0.5$ mm. The quantum zero-point electric field per photon $E_x$ along the gap direction is measured at the moir\'e plane, which is separated from the resonator by a 10 nm thick spacer. The dashed lines mark the lateral boundary of the metal structures.
 }
\label{fig-setup}
\end{figure*}

Nanoplasmonic split-ring resonators (SRR), with their deep-subwavelength confinement of terahertz or microwave photons, represent a remarkable ultrastrong coupling cavity platform~\cite{schlawin_cavity_2022,bayer_terahertz_2017,halbhuber_non-adiabatic_2020}. 
The cavity photon is primarily compressed into a submicron-to-micron-scale split gap, and the coupling to adjacent 2D electronic systems can exhibit a strength even exceeding the cavity resonance~\cite{bayer_terahertz_2017,halbhuber_non-adiabatic_2020}. 
These SRRs are also widely exploited as building blocks of metamaterials~\cite{pendry_magnetism_1999}, 
where multi-gap designs have been implemented to tailor negative permeability~\cite{penciu_multi-gap_2008}. 
Crucially, introducing multiple gaps on a ring preserves both the sharpness of the quantized resonances and the strength of the electric field within each gap.

Here we exploit the vacuum field of a multi-gap SRR for quantum manipulations of distributed quantum materials. In this architecture, a single cavity mode hosts spatially distributed field hot spots, each confined to a deep-subwavelength volume at a split gap, with hot spots separated by millimeter-scale distances.
The cavity vacuum fluctuation can thus mediate a many-body interaction to bridge the quantum materials distributed in the multiple split gaps into a single correlated mesoscopic system in equilibrium.
As a demonstration of this concept, we consider an SRR with two split gaps, each embedded with a hole-doped transition metal dichalcogenide (TMDs) moir\'{e} superlattice.
The cavity vacuum fluctuations mediate a current--current interaction between carriers in the two split gaps, whose sign depends on the relative current direction. This inherently gives rise to negative (attractive) channels, leading to remote Cooper pairing reminiscent of mesoscopic superconductivity, which we demonstrate using density matrix renormalization group (DMRG) and exact diagonalization (ED) calculations.
With the two constituents of a Cooper pair now remotely paired across a millimeter-scale separation, a hole incident at one split gap can be converted into an outgoing electron at the other gap, a process we term {\it Andreev teleportation}, which is demonstrated using the Green's function approach with a mean-field treatment of the vacuum-mediated interaction.
We find that the entanglement entropy between the two mesoscopic superlattices scales linearly with the total number of sites.







{\it Cavity-vacuum-mediated interaction between distributed moir\'e superlattices}---We consider a typical square-shaped metallic ring resonator with two split gaps separated by an arm length of 0.5 mm [Fig.~\ref{fig-setup}(a)]. Two TMD moir\'{e} superlattices are proximitized to the split gaps, with the vertical separation from the resonator plane set by the hBN spacer thickness. While TMD moir\'e superlattices can host a variety of quantum phases, here we start with the most common and experimentally accessible one---an otherwise trivial metal phase in a hole-doped heterobilayer described by a single-orbital triangular lattice (see Supplementary Note 1).
The two lowest-frequency cavity modes are shown in Figs.~\ref{fig-setup}(b) and \ref{fig-setup}(c), identified from finite-element simulations using COMSOL Multiphysics (c.f. Supplementary Note 6 for details). A single mode features two remotely separated electric field hot spots---one at each split gap---with deep-subwavelength confinement to a lateral region of $\sim 100 $ nm.
Moir\'e carriers within this spatial range are ultrastrongly coupled to the cavity photon, which is incorporated through the Peierls substitution,
\begin{equation}\label{eq-tb}
\hat H
=
-t \sum_{\langle ij\rangle\sigma}
e^{\,i\sum_{\lambda}\chi_{\lambda}
(\hat a_{\lambda}+\hat a_{\lambda}^{\dagger})
\xi_{\langle ij\rangle \sigma}^{\lambda}}
\hat c_{i\sigma}^{\dagger}\hat c_{j\sigma}
-\mu_0\sum_{i\sigma}\hat c_{i\sigma}^{\dagger}\hat c_{i\sigma}
+\sum_{\lambda}\hbar\omega_{\lambda}\hat a_{\lambda}^{\dagger}\hat a_{\lambda}.
\end{equation}
Here, $\sigma=A,B$ labels the two moir\'es. $\hat c_{i\sigma}^{\dagger}$ creates a hole at site $i$ in moir\'e $\sigma$, while $\hat a_{\lambda}^{\dagger}$ is photon creation operator of mode $\lambda$. We retain only nearest-neighbor hopping with amplitude $t=4.8$ meV fitted from miniband dispersion at moir\'e period of 6 nm  (Supplementary Fig. S1). Unless otherwise specified, energies below are measured in units of $t$.
We set the onsite energy of the two moir\'es to a common value $\mu_0$, while noting that they may in principle be different.
The geometric factor $\xi_{i j \sigma}^{\lambda} \equiv \boldsymbol{u}_{\lambda}\cdot \mathbf e_{\langle ij\rangle \sigma}$ encodes the dependence of the Peierls phase on the bond direction. $\mathbf e_{\langle ij\rangle \sigma}$ is the unit vector pointing from site $j$ to site $i$ in moir\'e $\sigma$, and $\boldsymbol{u}_{\lambda}$ is the polarization unit vector of the cavity field. The coupling strength $\chi_{\lambda}$ to a cavity mode is proportional to its vector-potential amplitude and thus scales as $\sqrt{1/\omega_{\lambda}}$.
See Supplementary Note 2 for details of the light--matter coupling.

The low-energy physics can be captured by expanding the Peierls phase to the first order and subsequently applying Schrieffer--Wolff transformation~\cite{schrieffer_relation_1966,cohen1998atom} to adiabatically eliminate the photonic degree of freedom (see Supplementary Note 3),
\begin{equation}\label{eq-effHam1}
\begin{aligned}
\hat{H}_{\mathrm{eff}}=&-t \sum_{\langle i j\rangle, \sigma} \hat{c}_{i, \sigma}^{\dagger} \hat{c}_{j, \sigma}-\mu_0 \sum_{i \sigma} \hat{c}_{i, \sigma}^{\dagger} \hat{c}_{i, \sigma}\\
&+\sum_{\lambda}\sum_{\langle i j\rangle } \sum_{\left\langle i^{\prime} j^{\prime}\right\rangle } V^{\lambda} \xi_{i j}^{\lambda} \xi_{i^{\prime} j^{\prime}}^{\lambda}\hat{c}_{i^{\prime},B}^{\dagger} \hat{c}_{i,A}^{\dagger} \hat{c}_{j,A} \hat{c}_{j^{\prime},B}.
\end{aligned}
\end{equation}
The last term is the cavity-vacuum-mediated carrier--carrier interaction, with a strength $V^{\lambda}=\frac{2 \chi_{\lambda}^2 t^2}{\hbar \omega_{\lambda}}\propto 1/\omega^2_{\lambda}$. This scaling implies the dominant role of the fundamental cavity mode---for which $V = 0.22$ for the simulated field in Fig.~\ref{fig-setup}(b)---while higher-frequency modes are expected to yield only modest quantitative corrections which we do not explicitly count.
This mediated interaction is of current--current type, with a sign that depends on the relative hopping directions of the two interacting holes. The negative interaction channels suggest a Cooper pairing instability. Within a split gap, this effect is overshadowed by the repulsive Coulomb interaction, which suppresses intra-gap pairing. Between different split gaps, however, Coulomb repulsion is completely negligible at the millimeter-scale separation. To limit the scope to its essential features, we retain only cavity-mediated interactions between holes in different split gaps.

By applying a unitary transformation that diagonalizes the single-particle terms, the low-energy effective Hamiltonian becomes (c.f. Supplementary Note 3):
\begin{equation}\label{eq-effHam2}
\begin{aligned}
&\hat{H}_{\mathrm{eff}}=\sum_{n,\sigma} \varepsilon_{n,\sigma}  \hat{\tilde{c}}_{n,\sigma}^{\dagger} \hat{\tilde{c}}_{n,\sigma}+\sum_{n n^{\prime} } \sum_{m m^{\prime}} \tilde{V}_{n  n^{\prime} m m^{\prime}} \hat{\tilde{c}}_{m,B}^{\dagger} \hat{\tilde{c}}_{n,A}^{\dagger} \hat{\tilde{c}}_{n^{\prime},A} \hat{\tilde{c}}_{m^{\prime},B}.\\
\end{aligned}
\end{equation}
 In the absence of cavity-mediated interaction, the system has a normal ground state by filling the single-particle levels in ascending order within each moir\'e superlattice.
The onset of cavity-vacuum-mediated interaction destabilizes this state as it scatters particle pair from quantum states $(n^{\prime},m^{\prime})$ to $(n,m)$ (Supplementary Fig. S1).
This process may stimulate pairing correlations in the ground state, characterized by the mixing of pairs across the Fermi level, analogous to the mechanism in mesoscopic superconductivity \cite{anderson_theory_1959,von_delft_parity-affected_1996, braun_superconductivity_1999}.
Specifically, it facilitates partial occupation pairs of states above the Fermi level while depleting certain pairs below it, thereby liberates the phase space for pair-scattering and ultimately lowering the ground-state energy.


\begin{figure}
\centering
\includegraphics[width=1\columnwidth]{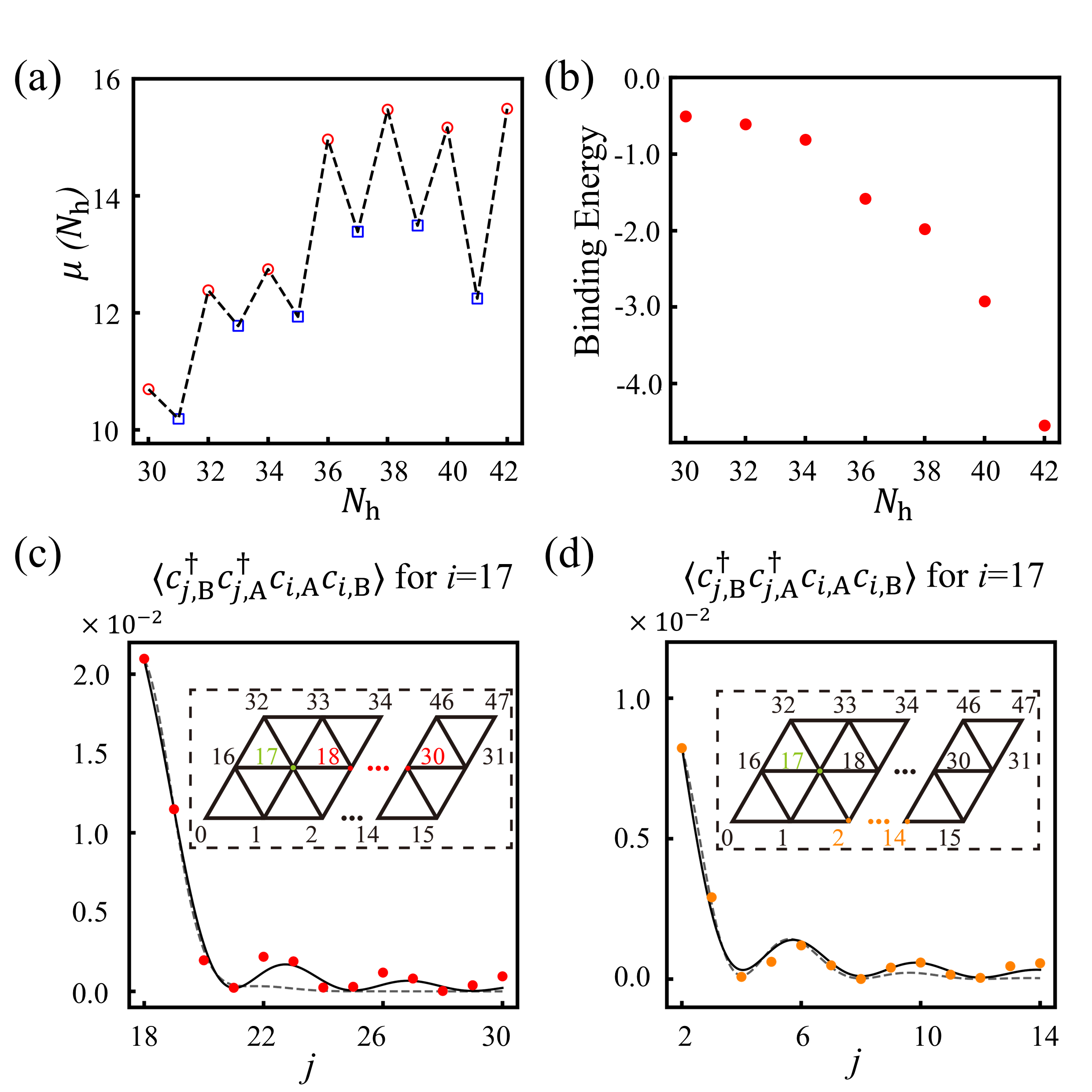}
\caption{\textbf{Many-body ground state under cavity-vacuum-mediated interaction---fixed-number DMRG calculations.}
\textbf{(a)} Chemical potential $\mu(N_h) \equiv E(N_h+1)-E(N_h)$ as a function of hole number $N_h$, where $E(N_h)$ is the ground-state energy of the two moir\'e superlattices (each of size  $6\times4$ sites). For even $N_h$, holes are equally partitioned ($N_h/2$ per superlattice). For odd $N_h$, moir\'e A has one more hole than moir\'e B.
\textbf{(b)} Binding energy defined as $\mu(N_h+1)-\mu(N_h)$ at even $N_h$.
\textbf{(c), (d)} Inter-moir\'e pair--pair correlation $\langle c^{\dagger}_{j,B}c^{\dagger}_{j,A} c_{i,A}c_{i,B} \rangle$, with $i$ being site $17$ (highlighted in green color) and $j$ varying along different trajectories (highlighted in red or orange color). Each moir\'e is of size $16\times3$ sites filled with 44 holes. These small system calculations use an interaction strength of $V=1$. Solid and dashed lines denote fits to $A x^{-\alpha}[1+c_1\cos(\pi x/2)]$ and $A\exp(-x/\xi)[1+c_1\cos(\pi x/2)]$, respectively. The power-law form provides better fit in both cases.}
\label{fig-dmrg}
\end{figure}

\begin{figure*}
\centering
\includegraphics[width=0.8\textwidth]{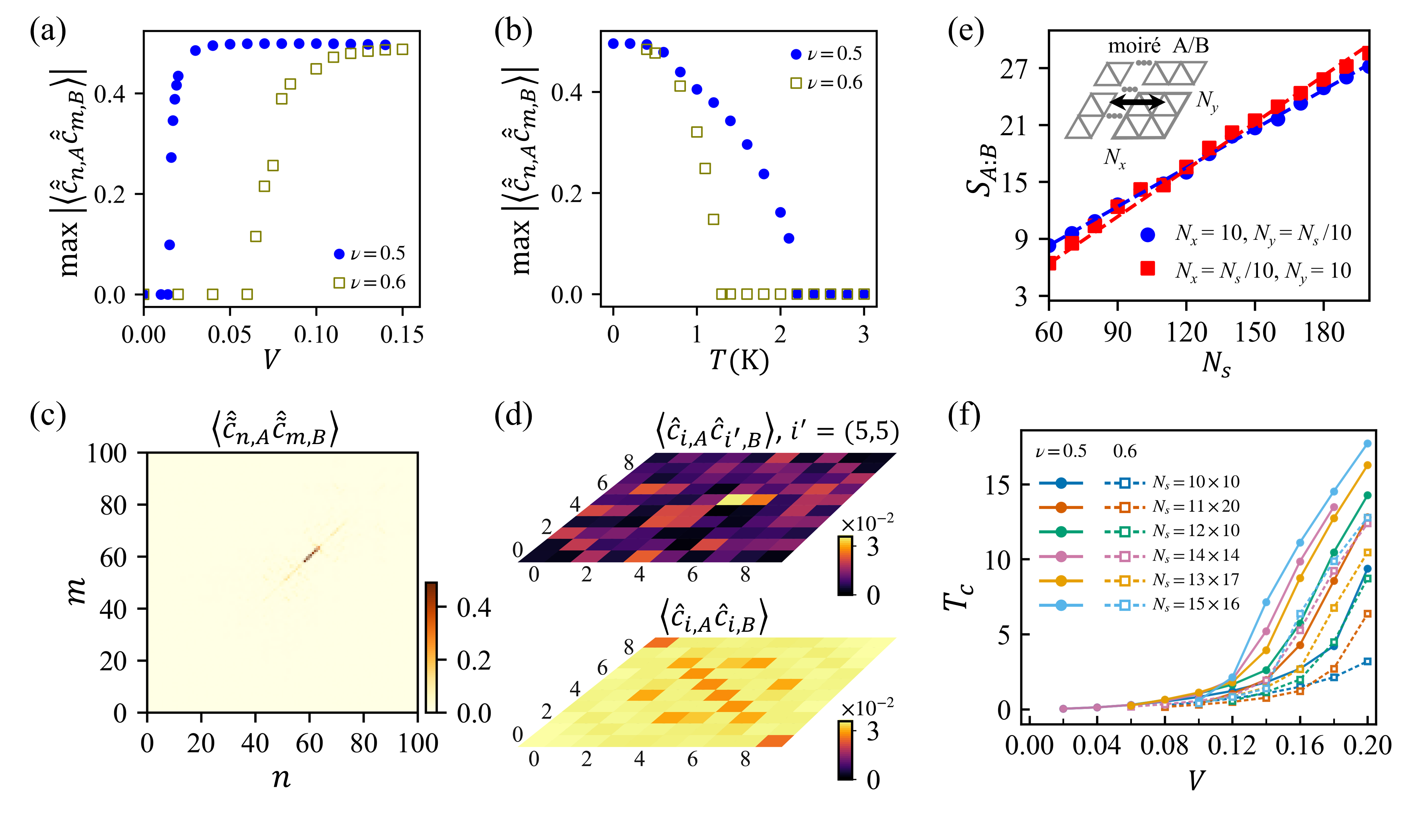}
\caption{\textbf{Remote pairing and distributed entanglement---mean-field solutions on larger-scale superlattices.}
\textbf{(a)} Maximum pairing amplitude $|\langle \hat{\tilde{c}}_{n,A}\hat{\tilde{c}}_{m,B} \rangle|$ at zero temperature, as function of the cavity-vacuum-mediated interaction $V$. Pairing emerges above a critical interaction strength that depends on the filling factor $\nu$.
\textbf{(b)} Temperature dependence of the maximum pairing amplitude at $V=0.15$, exhibiting a critical temperature $T_c$.
\textbf{(c), (d)} Pairing amplitude at zero temperature, in the single-particle eigenbasis (c), and in real space (d). The eigenbasis representation shows pronounced pairing near the Fermi level. In (a-d), each moir\'e is of $N_s= 10 \times 10$ sites. 
\textbf{(e)} Entanglement entropy between the two distributed mesoscopic moir\'es, as a function of their size $N_s$. The inset illustrates the lattice geometries corresponding to the red and blue symbols. In (c-e), $V=0.2$ and $\nu=0.6$. \textbf{(f)} Extracted $T_c$ at various moir\'e size $N_s$ and filling factor, as functions of the cavity-vacuum-mediated interaction strength $V$. The entire range of $V$ lies within that achievable with the COMSOL-simulated cavity mode in Fig.~\ref{fig-setup}(b), which realizes $V=0.22$ when the moir\'e plane is separated from the SRR by a 10 nm thick spacer.}
\label{fig-mf}
\end{figure*}

{\it Remote Cooper pairing}---To determine whether remote Cooper pairing is energetically favorable, we first employ real-space DMRG calculations using the Hamiltonian in Eq.~(\ref{eq-effHam1}) to compute the many-body ground-state energy $E(N_h)$ at fixed hole number $N_h$, and evaluate the binding energy $E(N_h)+E(N_h+2)-2E(N_h+1)$.
A negative binding energy signifies that the energy required to simultaneously  introduce two holes is less than twice the energy needed to introduce one hole, and therefore denotes the inclination to form pairing.
Figure~\ref{fig-dmrg}(a) shows the chemical potential $\mu(N_h) \equiv E(N_h+1)-E(N_h)$, i.e., the discrete addition energy for adding a single hole.
The even--odd oscillation already signifies the tendency of inter-split-gap pairing.
Figure~\ref{fig-dmrg}(b) plots the corresponding binding energy $\mu(N_h +1 ) - \mu(N_h)$, which is negative when the two moir\'es have equal number of holes.
The inter-moir\'e pair--pair correlation function, $\langle c^{\dagger}_{j,B}c^{\dagger}_{j,A} c_{i,A}c_{i,B} \rangle$, can be computed in the DMRG ground state. Figs.~\ref{fig-dmrg}(c) and \ref{fig-dmrg}(d) show the pair--pair correlations are better captured by a power-law decay with an oscillatory component, instead of an exponential form (Supplementary Note 8).

We also perform ED calculations on smaller systems. Benchmark calculations with the effective Hamiltonian Eq.~(\ref{eq-effHam1}) show excellent agreement with the DMRG results (Supplementary Note 8). We apply ED directly to the Hamiltonian in Eq.~(\ref{eq-tb}) with the photonic degree of freedom (Supplementary Note 7). The binding energy remains negative, providing further evidence for an energetic instability towards remote pairing between carriers across the two split gaps.

The DMRG and ED results establish a solid basis for introducing a mean-field treatment to access larger system with experimental relevance and to elucidate the underlying pairing mechanism.
Following the standard procedure of the mean-field approach to decouple the interaction term in the Hamiltonian specified in Eq. (\ref{eq-effHam2}), the ground state characterized by a pairing parameter $\langle \hat{\tilde{c}}_{n,A} \hat{\tilde{c}}_{m,B}  \rangle$ can be solved self-consistently (Supplementary Note 4).


The mean-field solutions are shown in Fig.~\ref{fig-mf}.
A finite inter-moir\'e pairing amplitude $\langle \hat{\tilde{c}}_{n,A} \hat{\tilde{c}}_{m,B} \rangle$ emerges above a critical interaction strength $V_{\rm c}$ (Fig.~\ref{fig-mf}(a)), accompanied by a drop in ground-state energy (Supplementary Fig.~S2), both evolving monotonically thereafter with $V$. Dependent on the filling, $V_{\rm c}$ can be one order of magnitude smaller than the value realized in the simulated cavity profile in Fig.~\ref{fig-setup}(b), leaving ample room for experimental realization.
The presence of pairing parameter in conjunction with the negative mean-field ground-state energy implies that the formation of remote Cooper pairs is energetically favorable.

The pairing parameter $\langle \hat{\tilde{c}}_{n,A} \hat{\tilde{c}}_{m,B} \rangle$ attains appreciable non-zero values primarily in the vicinity of the Fermi level, with the most pronounced contributions occurring for $m=n$ [Fig.~\ref{fig-mf}(c) and Supplementary Fig. S2], indicating preferential pairing between holes occupying {\it corresponding levels} in the two moir\'{e}s. This pairing structure is further confirmed by ED calculations of the full mesoscopic Hamiltonian in Eq.~(\ref{eq-tb}) (Supplementary Fig. S4). Such a pairing pattern corroborates the earlier hypothesis that cavity-mediated interactions induce a pair-correlated ground state, characterized by pairing mixing across the Fermi level, which is reminiscent of that in mesoscopic superconductivity \cite{anderson_theory_1959,von_delft_parity-affected_1996, braun_superconductivity_1999}.

In real space, the pairing amplitude $\langle \hat{c}_{i,A}\hat{c}_{j,B}\rangle$ exhibits a strong preference for {\it corresponding-site} pairing between the two moir\'{e}s  [Fig.~\ref{fig-mf}(d)]. This behavior can be understood from the structure of the cavity-vacuum-mediated interaction, which takes a current--current form with favored hopping directions selected by the cavity polarization. The resultant pairing therefore behaves effectively as a pair-hopping process favoring coherent nearest-neighbor motion of pairs. Corresponding-site inter-moir\'e pairing is energetically preferred as the available pair-hopping channels are maximized.

\begin{figure*}
\centering
\includegraphics[width=1\textwidth]{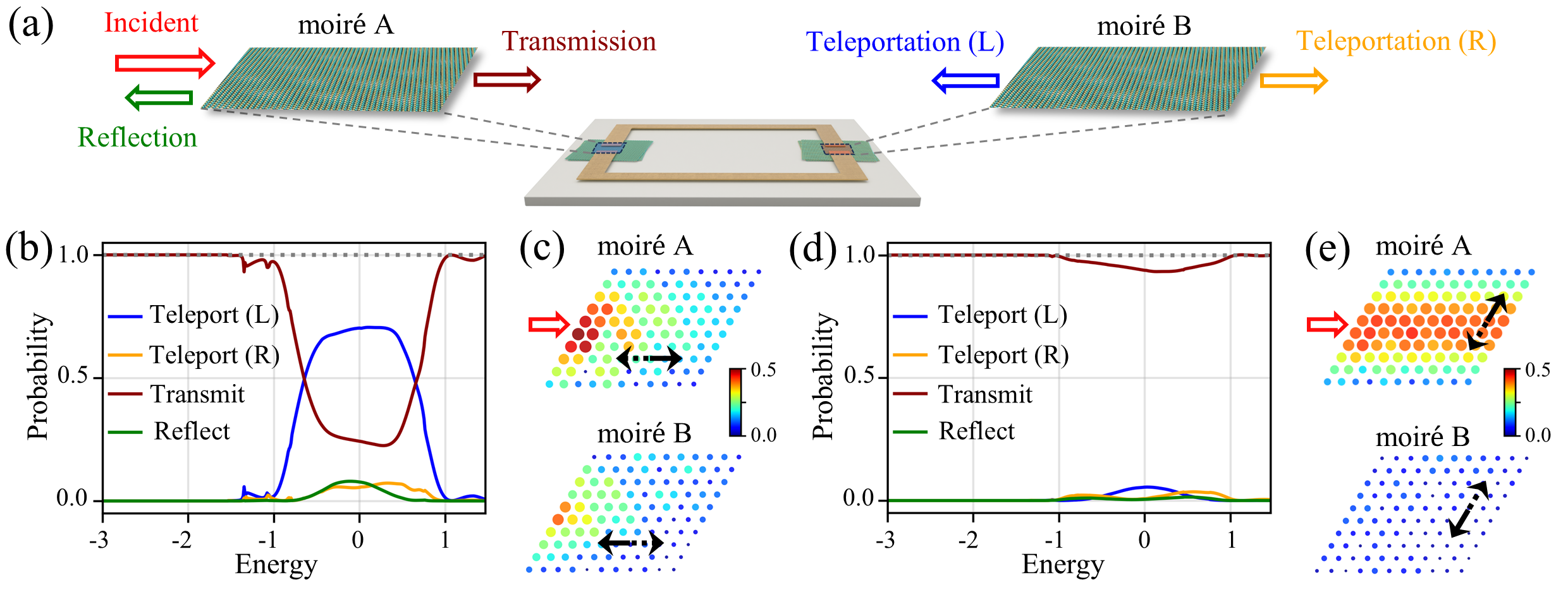}
\caption{\textbf{Andreev Teleportation.}
\textbf{(a)}
Schematic of the local scattering channels (normal reflection and transmission) and the remote Andreev teleportation channels (particle--hole interconversion across the two split gaps).
\textbf{(b)} Scattering probabilities as functions of incident energy, with the cavity field polarized along the incident direction. Inside the pairing gap, remote teleportation dominates over the local scattering channels. \textbf{(c)} The wavefunction of a scattering state at incident energy $E=0$, distributed across the moir\'{e} superlattices in both split gaps.
\textbf{(d), (e)} Similar plots for incidence with an angle of 60$^{\circ}$ relative to the cavity field polarization.
The calculations are performed with interaction strength $V=0.2$ and filling factor $\nu=0.6$, the same parameters used for the mean-field pairing maps in Figs.~\ref{fig-mf}(c) and \ref{fig-mf}(d); details are provided in Supplementary Note 9.
}
\label{fig_Andreev}
\end{figure*}


The mean-field results point to a BCS-like ground state in which each pair of corresponding levels (or corresponding sites in real space) from moir\'e A and moir\'e B respectively is in a coherent superposition of empty and pair-occupied configurations.
This implies significant bipartite entanglement $S_{A:B}$ between the two moir\'es. Indeed we find this entanglement in the mean-field ground state increases linearly with the total number of sites [Fig.~\ref{fig-mf}(e)]. The entanglement entropy associated with particle number fluctuations can be directly extracted from the pair-number distribution and contributes $\sim 30\%$ to $S_{A:B}$ (Supplementary Note 5).

From the temperature dependence (c.f. Fig.~\ref{fig-mf}(b)), we can extract the critical $T_c$ below which pairing occurs at the various moir\'e size $N_s$ and filling factor $\nu$. As shown in Fig.~\ref{fig-mf}(f), $T_c$ generally exhibits a trend to increase with $N_s$, while exhibiting quantum size effect with respect to both $N_s$ and $\nu$, reminiscent of the mesoscopic superconductivity~\cite{von_delft_superconductivity_2001,garcia-garcia_bcs_2011}. Larger density of states at the Fermi level leads to higher $T_c$.
The entire plot range is within that achievable with the COMSOL-simulated cavity mode in Fig.~\ref{fig-setup}(b), which realizes $V=0.22$ for moir\'e separated from the SRR by a 10 nm thick spacer. At this interaction strength, $T_c$ can reach nearly 20 K.


{\it Andreev teleportation}---Cooper-pair formation in a conventional superconductor gives rise to Andreev reflection: a hole from a normal metal enters the superconductor as part of a Cooper pair, while an electron is reflected back.
Here, Cooper-pair constituents are nonlocally paired across two distributed moir\'es. Upon a hole incident at moir\'e A, to form a Cooper pair for lowering the ground-state energy, the remote moir\'e B needs to draw a hole, thereby manifesting as an electron emission at moir\'e B.
Figure~\ref{fig_Andreev}(a) illustrates all possible scattering channels: local reflection/transmission within the same split gap, and Andreev teleportation to the remote split gap with particle--hole conversion.
To characterize these processes, we calculate the relevant scattering amplitudes within a Bogoliubov--de Gennes mode-matching framework \cite{ando_quantum_1991,khomyakov_conductance_2005}, formally equivalent to the Green's-function approach \cite{caroli_direct_1971,datta1995electronic}.
Figure~\ref{fig_Andreev}(b) presents the Andreev teleportation probability, as well as the local reflection and transmission probabilities, for cavity polarization along the direction of incidence [c.f. Fig.~\ref{fig_Andreev}(c)], using the same parameter set as in the mean-field calculations in Figs.~\ref{fig-mf}(c) and \ref{fig-mf}(d) (c.f. Supplementary Note 9).  Within the pairing gap, Andreev teleportation dominates over local scattering channels.
Figure~\ref{fig_Andreev}(c) shows the wavefunction of a scattering state at incident energy $E=0$. The profile explains that a hole injected into moir\'e A is predominantly converted into an electron emitting on the corresponding side of moir\'e B, as a manifestation of the corresponding-site inter-moir\'e pairing.

{\it Discussion and Outlook}---
We have demonstrated remote Cooper pairing between distributed mesoscopic lattices, made possible by cavity vacuum fluctuations with multiple hotspots in a multi-gap SRR. 
The ground state features large-scale distributed quantum entanglement, and enables Andreev teleportation over millimeter-scale distance.
This functionality is distinct from cross Andreev reflection concerning different leads in contact with a single piece of superconductor~\cite{feng_long-range_2025}, a single node geometry that also allows splitting of a Cooper pair into two non-interacting particles in different leads~\cite{hofstetter_cooper_2009}. The fundamental difference here is that particles from different nodes of the quantum network are interacting and paired in equilibrium, enabling nonlocal functionalities between the remotely separated nodes.

On the pairing nature, one notable observation is that the pair--pair correlation obtained from DMRG exhibits a power-law decay [Figs.~\ref{fig-dmrg}(c) and \ref{fig-dmrg}(d)], indicating quasi-long-range superconducting correlations. Its oscillatory modulation factor $1+c_1 \cos(\pi x/2)$, on the other hand, may suggest a finite-momentum pairing tendency.
The latter is corroborated by the incident direction dependence of the Andreev teleportation. As shown in Figs.~\ref{fig_Andreev}(d) and \ref{fig_Andreev}(e), when the incident direction spans a large angle with the cavity polarization, the teleportation probability drops significantly, from $78\%$ at $0^{\circ}$ to $9\%$ at $60^{\circ}$, which can be attributed to the reduction in available pair-hopping channels that can be mode matched to the incident hole. We also observe that above some interaction strength (which varies with $N_s$ and $\nu$ due to quantum-size effect), $T_c$ increases more rapidly with $V$, upon which the pairing amplitude also develops additional features (c.f. Supplementary Fig.~S6). 
These are all in line with an unconventional pairing nature, for instance, the possible emergence of intertwined superconducting-PDW order in this mesoscopic geometry, which warrants further investigation.

Finally, we would like to note that the multi-hotspot cavity vacuum could be generically realized by the evanescent field in micron or sub-micron scale split gaps, suitable for a variety of designs~\cite{appugliese_breakdown_2022,enkner_tunable_2025,xue_observation_2025}. The square-shaped SRR here is a straightforward example, which already provides sufficient vacuum fluctuation strength even without optimization. 
COMSOL simulation shows the hotspot spans a considerable range in the out-of-plane direction (Supplementary Fig.~S3), allowing the quantum materials of interest to be placed with ample vertical distance from the resonator plane. 
For example, using a typical 10-nm-thick hBN spacer between the moire and the SRR, the simulated vacuum field [Figs.~\ref{fig-setup}(b) and~\ref{fig-setup}(c)] can yield a mediated interaction strength that is one order of magnitude above the critical strength required [Fig.~\ref{fig-mf}(f)], while the critical temperature can reach nearly 20 K for this attainable interaction strength. The field hotspots define the mesoscopic regions where pairing occurs, seamlessly connected to the surrounding normal-metal region of the larger moir\'e for transport measurement access. 
All these characteristics point to a practically feasible experimental realization readily within reach.


{\it Acknowledgments}---We thank Shizhong Zhang, Zhongxia Shang and Cristiano Ciuti for helpful discussions. The work is supported by the National Natural Science Foundation of China (No. 12425406), Research Grant Council of Hong Kong SAR (A-HKU705/21, HKU SRFS2122-7S05, AoE/P-701/20), and New Cornerstone Science Foundation.

\bibliography{LM_SC}

@article{jarc_cavity-mediated_2023,
	title = {Cavity-mediated thermal control of metal-to-insulator transition in {1T}-{TaS2}},
	volume = {622},
	issn = {0028-0836, 1476-4687},
	url = {https://www.nature.com/articles/s41586-023-06596-2},
	doi = {10.1038/s41586-023-06596-2},
	number = {7983},
	urldate = {2024-03-06},
	journal = {Nature},
	author = {Jarc, Giacomo and Mathengattil, Shahla Yasmin and Montanaro, Angela and Giusti, Francesca and Rigoni, Enrico Maria and Sergo, Rudi and Fassioli, Francesca and Winnerl, Stephan and Dal Zilio, Simone and Mihailovic, Dragan and Prelov{\v{s}}ek, Peter and Eckstein, Martin and Fausti, Daniele},
	month = oct,
	year = {2023},
	pages = {487--492},
}

@article{chen_review_2016,
	title = {A review of metasurfaces: physics and applications},
	volume = {79},
	issn = {0034-4885, 1361-6633},
	shorttitle = {A review of metasurfaces},
	url = {https://iopscience.iop.org/article/10.1088/0034-4885/79/7/076401},
	doi = {10.1088/0034-4885/79/7/076401},
	number = {7},
	urldate = {2024-03-06},
	journal = {Rep. Prog. Phys.},
	author = {Chen, Hou-Tong and Taylor, Antoinette J and Yu, Nanfang},
	month = jul,
	year = {2016},
	pages = {076401},
}

@article{jeannin_ultrastrong_2019,
	title = {Ultrastrong {Light}--{Matter} {Coupling} in {Deeply} {Subwavelength} {THz} {LC} {Resonators}},
	volume = {6},
	issn = {2330-4022, 2330-4022},
	url = {https://pubs.acs.org/doi/10.1021/acsphotonics.8b01778},
	doi = {10.1021/acsphotonics.8b01778},
	number = {5},
	urldate = {2024-03-06},
	journal = {ACS Photonics},
	author = {Jeannin, Mathieu and Mariotti Nesurini, Giacomo and Suffit, St{\'e}phan and Gacemi, Djamal and Vasanelli, Angela and Li, Lianhe and Davies, Alexander Giles and Linfield, Edmund and Sirtori, Carlo and Todorov, Yanko},
	month = may,
	year = {2019},
	pages = {1207--1215},
}

@article{paravicini-bagliani_gate_2017,
	title = {Gate and magnetic field tunable ultrastrong coupling between a magnetoplasmon and the optical mode of an {LC} cavity},
	volume = {95},
	issn = {2469-9950, 2469-9969},
	url = {http://link.aps.org/doi/10.1103/PhysRevB.95.205304},
	doi = {10.1103/PhysRevB.95.205304},
	number = {20},
	urldate = {2024-03-06},
	journal = {Phys. Rev. B},
	author = {Paravicini-Bagliani, Gian L. and Scalari, Giacomo and Valmorra, Federico and Keller, Janine and Maissen, Curdin and Beck, Mattias and Faist, J{\'e}r{\^o}me},
	month = may,
	year = {2017},
	pages = {205304},
}

@article{paravicini-bagliani_magneto-transport_2019,
	title = {Magneto-transport controlled by {Landau} polariton states},
	volume = {15},
	issn = {1745-2473, 1745-2481},
	url = {https://www.nature.com/articles/s41567-018-0346-y},
	doi = {10.1038/s41567-018-0346-y},
	number = {2},
	urldate = {2024-03-06},
	journal = {Nature Phys},
	author = {Paravicini-Bagliani, Gian L. and Appugliese, Felice and Richter, Eli and Valmorra, Federico and Keller, Janine and Beck, Mattias and Bartolo, Nicola and R{\"o}ssler, Clemens and Ihn, Thomas and Ensslin, Klaus and Ciuti, Cristiano and Scalari, Giacomo and Faist, J{\'e}r{\^o}me},
	month = feb,
	year = {2019},
	pages = {186--190},
}

@article{scalari_ultrastrong_2012,
	title = {Ultrastrong {Coupling} of the {Cyclotron} {Transition} of a {2D} {Electron} {Gas} to a {THz} {Metamaterial}},
	volume = {335},
	issn = {0036-8075, 1095-9203},
	url = {https://www.science.org/doi/10.1126/science.1216022},
	doi = {10.1126/science.1216022},
	number = {6074},
	urldate = {2024-03-06},
	journal = {Science},
	author = {Scalari, G. and Maissen, C. and Tur{\v{c}}inkov{\'a}, D. and Hagenm{\"u}ller, D. and De Liberato, S. and Ciuti, C. and Reichl, C. and Schuh, D. and Wegscheider, W. and Beck, M. and Faist, J.},
	month = mar,
	year = {2012},
	pages = {1323--1326},
}

@article{appugliese_breakdown_2022,
	title = {Breakdown of topological protection by cavity vacuum fields in the integer quantum {Hall} effect},
	volume = {375},
	issn = {0036-8075, 1095-9203},
	url = {https://www.science.org/doi/10.1126/science.abl5818},
	doi = {10.1126/science.abl5818},
	number = {6584},
	urldate = {2024-03-06},
	journal = {Science},
	author = {Appugliese, Felice and Enkner, Josefine and Paravicini-Bagliani, Gian Lorenzo and Beck, Mattias and Reichl, Christian and Wegscheider, Werner and Scalari, Giacomo and Ciuti, Cristiano and Faist, J{\'e}r{\^o}me},
	month = mar,
	year = {2022},
	pages = {1030--1034},
}

@article{lin_remote_2023,
	title = {Remote gate control of topological transitions in moir{\'e} superlattices via cavity vacuum fields},
	volume = {120},
	issn = {0027-8424, 1091-6490},
	url = {https://pnas.org/doi/10.1073/pnas.2306584120},
	doi = {10.1073/pnas.2306584120},
	number = {32},
	urldate = {2024-05-17},
	journal = {Proc. Natl. Acad. Sci. U.S.A.},
	author = {Lin, Zuzhang and Xiao, Chengxin and Nguyen, Danh-Phuong and Arwas, Geva and Ciuti, Cristiano and Yao, Wang},
	month = aug,
	year = {2023},
	pages = {e2306584120},
}

@article{keller_few-electron_2017,
	title = {Few-{Electron} {Ultrastrong} {Light}-{Matter} {Coupling} at 300 {GHz} with {Nanogap} {Hybrid} {LC} {Microcavities}},
	volume = {17},
	issn = {1530-6984},
	url = {https://doi.org/10.1021/acs.nanolett.7b03228},
	doi = {10.1021/acs.nanolett.7b03228},
	number = {12},
	urldate = {2024-10-29},
	journal = {Nano Lett.},
	author = {Keller, Janine and Scalari, Giacomo and Cibella, Sara and Maissen, Curdin and Appugliese, Felice and Giovine, Ennio and Leoni, Roberto and Beck, Mattias and Faist, J{\'e}r{\^o}me},
	month = dec,
	year = {2017},
	note = {Publisher: American Chemical Society},
	pages = {7410--7415},
}

@article{sentef_cavity_2018,
	title = {Cavity quantum-electrodynamical polaritonically enhanced electron-phonon coupling and its influence on superconductivity},
	volume = {4},
	url = {https://www.science.org/doi/10.1126/sciadv.aau6969},
	doi = {10.1126/sciadv.aau6969},
	number = {11},
	urldate = {2025-02-17},
	journal = {Sci. Adv.},
	publisher = {American Association for the Advancement of Science},
	author = {Sentef, M. A. and Ruggenthaler, M. and Rubio, A.},
	month = nov,
	year = {2018},
	pages = {eaau6969}
}

@article{schlawin2019cavity,
	title = {Cavity-{Mediated} {Electron}-{Photon} {Superconductivity}},
	volume = {122},
	issn = {0031-9007, 1079-7114},
	url = {https://link.aps.org/doi/10.1103/PhysRevLett.122.133602},
	doi = {10.1103/PhysRevLett.122.133602},
	number = {13},
	urldate = {2024-03-06},
	journal = {Phys. Rev. Lett.},
	author = {Schlawin, Frank and Cavalleri, Andrea and Jaksch, Dieter},
	month = apr,
	year = {2019},
	pages = {133602}
}

@article{andolina_amperean_2024,
	title = {Amperean superconductivity cannot be induced by deep subwavelength cavities in a two-dimensional material},
	volume = {109},
	url = {https://link.aps.org/doi/10.1103/PhysRevB.109.104513},
	doi = {10.1103/PhysRevB.109.104513},
	number = {10},
	journal = {Phys. Rev. B},
	publisher = {American Physical Society},
	author = {Andolina, Gian Marcello and De Pasquale, Antonella and Pellegrino, Francesco Maria Dimitri and Torre, Iacopo and Koppens, Frank H. L. and Polini, Marco},
	month = mar,
	year = {2024},
	pages = {104513}
}

@article{gao2020photoinduced,
	title = {Photoinduced {Electron} {Pairing} in a {Driven} {Cavity}},
	volume = {125},
	issn = {0031-9007, 1079-7114},
	url = {https://link.aps.org/doi/10.1103/PhysRevLett.125.053602},
	doi = {10.1103/PhysRevLett.125.053602},
	number = {5},
	urldate = {2024-03-06},
	journal = {Phys. Rev. Lett.},
	author = {Gao, Hongmin and Schlawin, Frank and Buzzi, Michele and Cavalleri, Andrea and Jaksch, Dieter},
	month = jul,
	year = {2020},
	pages = {053602}
}

@article{kozin_cavity-enhanced_2025,
	title = {Cavity-enhanced superconductivity via band engineering},
	volume = {111},
	url = {https://link.aps.org/doi/10.1103/PhysRevB.111.035410},
	doi = {10.1103/PhysRevB.111.035410},
	number = {3},
	urldate = {2026-01-23},
	journal = {Phys. Rev. B},
	publisher = {American Physical Society},
	author = {Kozin, Valerii K. and Thingstad, Even and Loss, Daniel and Klinovaja, Jelena},
	month = jan,
	year = {2025},
	pages = {035410}
}

@article{li2020manipulating,
  title={Manipulating intertwined orders in solids with quantum light},
  author={Li, Jiajun and Eckstein, Martin},
  journal={Phys. Rev. Lett.},
  volume={125},
  number={21},
  pages={217402},
  year={2020},
  publisher={APS}
}

@article{anderson_theory_1959,
	title = {Theory of dirty superconductors},
	volume = {11},
	issn = {0022-3697},
	url = {https://www.sciencedirect.com/science/article/pii/0022369759900368},
	doi = {10.1016/0022-3697(59)90036-8},
	number = {1},
	urldate = {2025-10-17},
	journal = {J. Phys. Chem. Solids},
	author = {Anderson, P. W.},
	month = sep,
	year = {1959},
	pages = {26--30},
}

@article{von_delft_parity-affected_1996,
	title = {Parity-{Affected} {Superconductivity} in {Ultrasmall} {Metallic} {Grains}},
	volume = {77},
	url = {https://link.aps.org/doi/10.1103/PhysRevLett.77.3189},
	doi = {10.1103/PhysRevLett.77.3189},
	number = {15},
	urldate = {2024-12-16},
	journal = {Phys. Rev. Lett.},
	author = {von Delft, Jan and Zaikin, Andrei D. and Golubev, Dmitrii S. and Tichy, Wolfgang},
	month = oct,
	year = {1996},
	pages = {3189--3192},
}

@article{braun_superconductivity_1999,
	title = {Superconductivity in ultrasmall metallic grains},
	volume = {59},
	url = {https://link.aps.org/doi/10.1103/PhysRevB.59.9527},
	doi = {10.1103/PhysRevB.59.9527},
	number = {14},
	urldate = {2024-12-12},
	journal = {Phys. Rev. B},
	author = {Braun, Fabian and von Delft, Jan},
	month = apr,
	year = {1999},
	pages = {9527--9544},
}

@article{enkner_tunable_2025,
	title = {Tunable vacuum-field control of fractional and integer quantum {Hall} phases},
	volume = {641},
	copyright = {2025 The Author(s)},
	issn = {1476-4687},
	url = {https://www.nature.com/articles/s41586-025-08894-3},
	doi = {10.1038/s41586-025-08894-3},
	number = {8064},
	urldate = {2026-01-18},
	journal = {Nature},
	publisher = {Nature Publishing Group},
	author = {Enkner, Josefine and Graziotto, Lorenzo and Bori{\c{c}}i, Dalin and Appugliese, Felice and Reichl, Christian and Scalari, Giacomo and Regnault, Nicolas and Wegscheider, Werner and Ciuti, Cristiano and Faist, J{\'e}r{\^o}me},
	month = may,
	year = {2025},
	pages = {884--889}
}

@article{thomas_tilting_2019,
	title = {Tilting a ground-state reactivity landscape by vibrational strong coupling},
	volume = {363},
	url = {https://www.science.org/doi/full/10.1126/science.aau7742},
	doi = {10.1126/science.aau7742},
	number = {6427},
	urldate = {2024-04-04},
	journal = {Science},
	author = {Thomas, A. and Lethuillier-Karl, L. and Nagarajan, K. and Vergauwe, R. M. A. and George, J. and Chervy, T. and Shalabney, A. and Devaux, E. and Genet, C. and Moran, J. and Ebbesen, T. W.},
	month = feb,
	year = {2019},
	pages = {615--619}
}

@article{frisk_kockum_ultrastrong_2019,
	title = {Ultrastrong coupling between light and matter},
	volume = {1},
	issn = {2522-5820},
	url = {https://www.nature.com/articles/s42254-018-0006-2},
	doi = {10.1038/s42254-018-0006-2},
	number = {1},
	urldate = {2024-03-06},
	journal = {Nat. Rev. Phys.},
	author = {Frisk Kockum, Anton and Miranowicz, Adam and De Liberato, Simone and Savasta, Salvatore and Nori, Franco},
	month = jan,
	year = {2019},
	pages = {19--40}
}

@article{garcia-vidal_manipulating_2021,
	title = {Manipulating matter by strong coupling to vacuum fields},
	volume = {373},
	issn = {0036-8075, 1095-9203},
	url = {https://www.science.org/doi/10.1126/science.abd0336},
	doi = {10.1126/science.abd0336},
	number = {6551},
	urldate = {2024-03-06},
	journal = {Science},
	author = {Garcia-Vidal, Francisco J. and Ciuti, Cristiano and Ebbesen, Thomas W.},
	month = jul,
	year = {2021},
	pages = {eabd0336}
}

@article{schlawin_cavity_2022,
	title = {Cavity quantum materials},
	volume = {9},
	issn = {1931-9401},
	url = {https://doi.org/10.1063/5.0083825},
	doi = {10.1063/5.0083825},
	number = {1},
	urldate = {2024-04-26},
	journal = {Appl. Phys. Rev.},
	author = {Schlawin, F. and Kennes, D. M. and Sentef, M. A.},
	month = feb,
	year = {2022},
	pages = {011312}
}

@article{forn-diaz_ultrastrong_2019,
	title = {Ultrastrong coupling regimes of light-matter interaction},
	volume = {91},
	issn = {0034-6861, 1539-0756},
	url = {https://link.aps.org/doi/10.1103/RevModPhys.91.025005},
	doi = {10.1103/RevModPhys.91.025005},
	number = {2},
	urldate = {2024-03-06},
	journal = {Rev. Mod. Phys.},
	author = {Forn-D{\'i}az, P. and Lamata, L. and Rico, E. and Kono, J. and Solano, E.},
	month = jun,
	year = {2019},
	pages = {025005}
}

@article{lu_cavity_2025,
	title = {Cavity engineering of solid-state materials without external driving},
	volume = {17},
	issn = {1943-8206},
	url = {https://opg.optica.org/abstract.cfm?URI=aop-17-2-441},
	doi = {10.1364/AOP.544138},
	number = {2},
	urldate = {2025-09-20},
	journal = {Adv. Opt. Photonics},
	author = {Lu, I-Te and Shin, Dongbin and Kamper Svendsen, Mark and Latini, Simone and H{\"u}bener, Hannes and Ruggenthaler, Michael and Rubio, Angel},
	month = jun,
	year = {2025},
	pages = {441}
}

@article{wei_cavity-vacuum-induced_2025,
	title = {Cavity-{Vacuum}-{Induced} {Chiral} {Spin} {Liquids} in {Kagome} {Lattices}: {Tuning} and {Probing} {Topological} {Quantum} {Phases} via {Cavity} {Quantum} {Electrodynamics}},
	volume = {135},
	shorttitle = {Cavity-{Vacuum}-{Induced} {Chiral} {Spin} {Liquids} in {Kagome} {Lattices}},
	url = {https://link.aps.org/doi/10.1103/8qx2-xxh2},
	doi = {10.1103/8qx2-xxh2},
	number = {23},
	urldate = {2026-01-18},
	journal = {Phys. Rev. Lett.},
	publisher = {American Physical Society},
	author = {Wei, Chenan and Yang, Liu and Jiang, Qing-Dong},
	month = dec,
	year = {2025},
	pages = {236901},
}

@article{lu_cavity-enhanced_2024,
	title = {Cavity-enhanced superconductivity in {MgB}$_{\textrm{2}}$ from first-principles quantum electrodynamics ({QEDFT})},
	volume = {121},
	issn = {0027-8424, 1091-6490},
	url = {https://pnas.org/doi/10.1073/pnas.2415061121},
	doi = {10.1073/pnas.2415061121},
	number = {50},
	urldate = {2026-01-21},
	journal = {Proc. Natl. Acad. Sci. U. S. A.},
	author = {Lu, I-Te and Shin, Dongbin and Svendsen, Mark Kamper and H{\"u}bener, Hannes and De Giovannini, Umberto and Latini, Simone and Ruggenthaler, Michael and Rubio, Angel},
	month = dec,
	year = {2024},
	pages = {e2415061121}
}

@article{mivehvar_cavity_2021,
	title = {Cavity {QED} with quantum gases: new paradigms in many-body physics},
	volume = {70},
	issn = {0001-8732},
	shorttitle = {Cavity {QED} with quantum gases},
	url = {https://doi.org/10.1080/00018732.2021.1969727},
	doi = {10.1080/00018732.2021.1969727},
	number = {1},
	urldate = {2026-02-26},
	journal = {Adv. Phys.},
	publisher = {Taylor \& Francis},
	author = {Mivehvar, Farokh and Piazza, Francesco and Donner, Tobias and Ritsch, Helmut},
	month = jan,
	year = {2021},
	pages = {1--153},
}

@article{chang_colloquium_2018,
	title = {Colloquium: {Quantum} matter built from nanoscopic lattices of atoms and photons},
	volume = {90},
	shorttitle = {Colloquium},
	url = {https://link.aps.org/doi/10.1103/RevModPhys.90.031002},
	doi = {10.1103/RevModPhys.90.031002},
	number = {3},
	urldate = {2026-03-13},
	journal = {Reviews of Modern Physics},
	publisher = {American Physical Society},
	author = {Chang, D.E. and Douglas, J.S. and Gonz{\'a}lez-Tudela, A. and Hung, C.-L. and Kimble, H.J.},
	month = aug,
	year = {2018},
	pages = {031002},
}

@article{blais_circuit_2021,
	title = {Circuit quantum electrodynamics},
	volume = {93},
	url = {https://link.aps.org/doi/10.1103/RevModPhys.93.025005},
	doi = {10.1103/RevModPhys.93.025005},
	number = {2},
	urldate = {2026-03-13},
	journal = {Reviews of Modern Physics},
	publisher = {American Physical Society},
	author = {Blais, Alexandre and Grimsmo, Arne L. and Girvin, S.M. and Wallraff, Andreas},
	month = may,
	year = {2021},
	pages = {025005},
}

@article{ritsch_cold_2013,
	title = {Cold atoms in cavity-generated dynamical optical potentials},
	volume = {85},
	url = {https://link.aps.org/doi/10.1103/RevModPhys.85.553},
	doi = {10.1103/RevModPhys.85.553},
	number = {2},
	urldate = {2026-03-13},
	journal = {Reviews of Modern Physics},
	publisher = {American Physical Society},
	author = {Ritsch, Helmut and Domokos, Peter and Brennecke, Ferdinand and Esslinger, Tilman},
	month = apr,
	year = {2013},
	pages = {553--601},
}

@article{pellizzari_decoherence_1995,
	title = {Decoherence, {Continuous} {Observation}, and {Quantum} {Computing}: {A} {Cavity} {QED} {Model}},
	volume = {75},
	shorttitle = {Decoherence, {Continuous} {Observation}, and {Quantum} {Computing}},
	url = {https://link.aps.org/doi/10.1103/PhysRevLett.75.3788},
	doi = {10.1103/PhysRevLett.75.3788},
	number = {21},
	urldate = {2026-03-13},
	journal = {Physical Review Letters},
	publisher = {American Physical Society},
	author = {Pellizzari, T. and Gardiner, S. A. and Cirac, J. I. and Zoller, P.},
	month = nov,
	year = {1995},
	pages = {3788--3791}
}

@article{imamoglu_quantum_1999,
	title = {Quantum {Information} {Processing} {Using} {Quantum} {Dot} {Spins} and {Cavity} {QED}},
	volume = {83},
	url = {https://link.aps.org/doi/10.1103/PhysRevLett.83.4204},
	doi = {10.1103/PhysRevLett.83.4204},
	number = {20},
	urldate = {2026-03-13},
	journal = {Physical Review Letters},
	publisher = {American Physical Society},
	author = {{\.I}mamo{\u{g}}lu, A. and Awschalom, D. D. and Burkard, G. and DiVincenzo, D. P. and Loss, D. and Sherwin, M. and Small, A.},
	month = nov,
	year = {1999},
	pages = {4204--4207}
}

@article{dimer_proposed_2007,
	title = {Proposed realization of the {Dicke}-model quantum phase transition in an optical cavity {QED} system},
	volume = {75},
	url = {https://link.aps.org/doi/10.1103/PhysRevA.75.013804},
	doi = {10.1103/PhysRevA.75.013804},
	number = {1},
	urldate = {2026-03-13},
	journal = {Physical Review A},
	publisher = {American Physical Society},
	author = {Dimer, F. and Estienne, B. and Parkins, A. S. and Carmichael, H. J.},
	month = jan,
	year = {2007},
	pages = {013804}
}

@article{baumann_dicke_2010,
	title = {Dicke quantum phase transition with a superfluid gas in an optical cavity},
	volume = {464},
	copyright = {2010 Macmillan Publishers Limited. All rights reserved},
	issn = {1476-4687},
	url = {https://www.nature.com/articles/nature09009},
	doi = {10.1038/nature09009},
	number = {7293},
	urldate = {2026-03-13},
	journal = {Nature},
	publisher = {Nature Publishing Group},
	author = {Baumann, Kristian and Guerlin, Christine and Brennecke, Ferdinand and Esslinger, Tilman},
	month = apr,
	year = {2010},
	pages = {1301--1306}
}

@article{cirac_quantum_1997,
	title = {Quantum {State} {Transfer} and {Entanglement} {Distribution} among {Distant} {Nodes} in a {Quantum} {Network}},
	volume = {78},
	url = {https://link.aps.org/doi/10.1103/PhysRevLett.78.3221},
	doi = {10.1103/PhysRevLett.78.3221},
	number = {16},
	urldate = {2026-03-14},
	journal = {Physical Review Letters},
	publisher = {American Physical Society},
	author = {Cirac, J. I. and Zoller, P. and Kimble, H. J. and Mabuchi, H.},
	month = apr,
	year = {1997},
	pages = {3221--3224}
}

@article{yao_theory_2005,
	title = {Theory of {Control} of the {Spin}-{Photon} {Interface} for {Quantum} {Networks}},
	volume = {95},
	url = {https://link.aps.org/doi/10.1103/PhysRevLett.95.030504},
	doi = {10.1103/PhysRevLett.95.030504},
	number = {3},
	urldate = {2026-03-14},
	journal = {Physical Review Letters},
	publisher = {American Physical Society},
	author = {Yao, Wang and Liu, Ren-Bao and Sham, L. J.},
	month = jul,
	year = {2005},
	pages = {030504}
}

@article{duan_long-distance_2001,
	title = {Long-distance quantum communication with atomic ensembles and linear optics},
	volume = {414},
	copyright = {2001 Macmillan Magazines Ltd.},
	issn = {1476-4687},
	url = {https://www.nature.com/articles/35106500},
	doi = {10.1038/35106500},
	number = {6862},
	urldate = {2026-03-14},
	journal = {Nature},
	publisher = {Nature Publishing Group},
	author = {Duan, L.-M. and Lukin, M. D. and Cirac, J. I. and Zoller, P.},
	month = nov,
	year = {2001},
	pages = {413--418}
}

@article{bloch_strongly_2022,
	title = {Strongly correlated electron--photon systems},
	volume = {606},
	copyright = {2022 This is a U.S. government work and not under copyright protection in the U.S.; foreign copyright protection may apply},
	issn = {1476-4687},
	url = {https://www.nature.com/articles/s41586-022-04726-w},
	doi = {10.1038/s41586-022-04726-w},
	number = {7912},
	urldate = {2026-03-14},
	journal = {Nature},
	publisher = {Nature Publishing Group},
	author = {Bloch, Jacqueline and Cavalleri, Andrea and Galitski, Victor and Hafezi, Mohammad and Rubio, Angel},
	month = jun,
	year = {2022},
	pages = {41--48}
}

@article{thomas_exploring_2025,
	title = {Exploring superconductivity under strong coupling with the vacuum electromagnetic field},
	volume = {162},
	issn = {0021-9606},
	url = {https://doi.org/10.1063/5.0231202},
	doi = {10.1063/5.0231202},
	number = {13},
	urldate = {2026-03-14},
	journal = {The Journal of Chemical Physics},
	author = {Thomas, A. and Devaux, E. and Nagarajan, K. and Chervy, T. and Seidel, M. and Rogez, G. and Robert, J. and Drillon, M. and Ruan, T. T. and Schlittenhardt, S. and Ruben, M. and Hagenm{\"u}ller, D. and Sch{\"u}tz, S. and Schachenmayer, J. and Genet, C. and Pupillo, G. and Ebbesen, T. W.},
	month = apr,
	year = {2025},
	pages = {134701}
}

@article{ashida_quantum_2020,
	title = {Quantum {Electrodynamic} {Control} of {Matter}: {Cavity}-{Enhanced} {Ferroelectric} {Phase} {Transition}},
	volume = {10},
	issn = {2160-3308},
	shorttitle = {Quantum {Electrodynamic} {Control} of {Matter}},
	url = {https://link.aps.org/doi/10.1103/PhysRevX.10.041027},
	doi = {10.1103/PhysRevX.10.041027},
	number = {4},
	urldate = {2024-03-06},
	journal = {Physical Review X},
	author = {Ashida, Yuto and {\.I}mamo{\u{g}}lu, Ata{\c{c}} and Faist, J{\'e}r{\^o}me and Jaksch, Dieter and Cavalleri, Andrea and Demler, Eugene},
	month = nov,
	year = {2020},
	pages = {041027}
}

@article{jiang_quantum_2019,
	title = {Quantum atmospherics for materials diagnosis},
	volume = {99},
	url = {https://link.aps.org/doi/10.1103/PhysRevB.99.201104},
	doi = {10.1103/PhysRevB.99.201104},
	number = {20},
	journal = {Phys. Rev. B},
	author = {Jiang, Qing-Dong and Wilczek, Frank},
	month = may,
	year = {2019},
	pages = {201104}
}

@article{hubener_engineering_2021,
	title = {Engineering quantum materials with chiral optical cavities},
	volume = {20},
	issn = {1476-1122, 1476-4660},
	url = {https://www.nature.com/articles/s41563-020-00801-7},
	doi = {10.1038/s41563-020-00801-7},
	number = {4},
	urldate = {2024-03-06},
	journal = {Nature Materials},
	author = {H{\"u}bener, Hannes and De Giovannini, Umberto and Sch{\"a}fer, Christian and Andberger, Johan and Ruggenthaler, Michael and Faist, Jerome and Rubio, Angel},
	month = apr,
	year = {2021},
	pages = {438--442}
}

@article{ke_vacuum-induced_2023,
	title = {Vacuum-{Induced} {Symmetry} {Breaking} of {Chiral} {Enantiomer} {Formation} in {Chemical} {Reactions}},
	volume = {131},
	issn = {0031-9007, 1079-7114},
	url = {https://link.aps.org/doi/10.1103/PhysRevLett.131.223601},
	doi = {10.1103/PhysRevLett.131.223601},
	number = {22},
	urldate = {2024-03-06},
	journal = {Physical Review Letters},
	author = {Ke, Yanzhe and Song, Zhigang and Jiang, Qing-Dong},
	month = nov,
	year = {2023},
	pages = {223601}
}

@article{lin_spontaneous_2026,
	title = {Spontaneous {Symmetry} {Breaking} of {Cavity} {Vacuum} and {Emergent} {Gyrotropic} {Effects} in {Embedded} {Moir}{\textbackslash}'e {Superlattices}},
	volume = {136},
	url = {https://link.aps.org/doi/10.1103/1cgv-x7tm},
	doi = {10.1103/1cgv-x7tm},
	number = {4},
	urldate = {2026-02-01},
	journal = {Physical Review Letters},
	publisher = {American Physical Society},
	author = {Lin, Zuzhang and Chan, Hsun-Chi and Yang, Wenqi and Sha, Yixin and Xiao, Cong and Zhang, Shuang and Yao, Wang},
	month = jan,
	year = {2026},
	pages = {046903}
}

@article{curtis_cavity_2019,
	title = {Cavity {Quantum} {Eliashberg} {Enhancement} of {Superconductivity}},
	volume = {122},
	issn = {0031-9007, 1079-7114},
	url = {https://link.aps.org/doi/10.1103/PhysRevLett.122.167002},
	doi = {10.1103/PhysRevLett.122.167002},
	number = {16},
	urldate = {2024-03-06},
	journal = {Physical Review Letters},
	author = {Curtis, Jonathan B. and Raines, Zachary M. and Allocca, Andrew A. and Hafezi, Mohammad and Galitski, Victor M.},
	month = apr,
	year = {2019},
	pages = {167002}
}

@article{bayer_terahertz_2017,
	title = {Terahertz {Light}--{Matter} {Interaction} beyond {Unity} {Coupling} {Strength}},
	volume = {17},
	issn = {1530-6984, 1530-6992},
	url = {https://pubs.acs.org/doi/10.1021/acs.nanolett.7b03103},
	doi = {10.1021/acs.nanolett.7b03103},
	number = {10},
	urldate = {2024-03-06},
	journal = {Nano Letters},
	author = {Bayer, Andreas and Pozimski, Marcel and Schambeck, Simon and Schuh, Dieter and Huber, Rupert and Bougeard, Dominique and Lange, Christoph},
	month = oct,
	year = {2017},
	pages = {6340--6344},
}

@article{halbhuber_non-adiabatic_2020,
	title = {Non-adiabatic stripping of a cavity field from electrons in the deep-strong coupling regime},
	volume = {14},
	copyright = {2020 The Author(s), under exclusive licence to Springer Nature Limited},
	issn = {1749-4893},
	url = {https://www.nature.com/articles/s41566-020-0673-2},
	doi = {10.1038/s41566-020-0673-2},
	number = {11},
	urldate = {2026-03-14},
	journal = {Nature Photonics},
	publisher = {Nature Publishing Group},
	author = {Halbhuber, M. and Mornhinweg, J. and Zeller, V. and Ciuti, C. and Bougeard, D. and Huber, R. and Lange, C.},
	month = nov,
	year = {2020},
	pages = {675--679},
}

@article{pendry_magnetism_1999,
	title = {Magnetism from conductors and enhanced nonlinear phenomena},
	volume = {47},
	issn = {1557-9670},
	url = {https://ieeexplore.ieee.org/document/798002},
	doi = {10.1109/22.798002},
	number = {11},
	urldate = {2026-03-14},
	journal = {IEEE Transactions on Microwave Theory and Techniques},
	author = {Pendry, J.B. and Holden, A.J. and Robbins, D.J. and Stewart, W.J.},
	month = nov,
	year = {1999},
	pages = {2075--2084}
}

@article{penciu_multi-gap_2008,
	title = {Multi-gap individual and coupled split-ring resonator structures},
	volume = {16},
	issn = {1094-4087},
	url = {https://opg.optica.org/oe/abstract.cfm?uri=oe-16-22-18131},
	doi = {10.1364/OE.16.018131},
	number = {22},
	urldate = {2024-03-06},
	journal = {Optics Express},
	author = {Penciu, R. S. and Aydin, K. and Kafesaki, M. and Koschny, Th. and Ozbay, E. and Economou, E. N. and Soukoulis, C. M.},
	month = oct,
	year = {2008},
	pages = {18131}
}

@article{caroli_direct_1971,
	title = {Direct calculation of the tunneling current},
	volume = {4},
	issn = {0022-3719},
	url = {https://doi.org/10.1088/0022-3719/4/8/018},
	doi = {10.1088/0022-3719/4/8/018},
	number = {8},
	urldate = {2026-03-18},
	journal = {Journal of Physics C: Solid State Physics},
	author = {Caroli, C. and Combescot, R. and Nozieres, P. and Saint-James, D.},
	month = jun,
	year = {1971},
	pages = {916}
}

@book{datta1995electronic,
  title     = {Electronic Transport in Mesoscopic Systems},
  author    = {Datta, Supriyo},
  year      = {1995},
  publisher = {Cambridge University Press},
  address   = {Cambridge, England},
  isbn      = {9780521416047},
  series    = {Cambridge Studies in Semiconductor Physics and Microelectronic Engineering},
}

@article{khomyakov_conductance_2005,
	title = {Conductance calculations for quantum wires and interfaces: Mode matching and Green's functions},
	volume = {72},
	issn = {1098-0121},
	url = {https://doi.org/10.1103/PhysRevB.72.035450},
	doi = {10.1103/PhysRevB.72.035450},
	number = {3},
	urldate = {2026-03-18},
	journal = {Phys. Rev. B},
	author = {Khomyakov, P. A. and Brocks, G. and Karpan, V. and Zwierzycki, M. and Kelly, P. J.},
	month = jul,
	year = {2005},
	pages = {035450}
}

@article{ando_quantum_1991,
	title = {Quantum point contacts in magnetic fields},
	volume = {44},
	issn = {0163-1829},
	url = {https://doi.org/10.1103/PhysRevB.44.8017},
	doi = {10.1103/PhysRevB.44.8017},
	number = {15},
	urldate = {2026-03-18},
	journal = {Phys. Rev. B},
	author = {Ando, T.},
	month = oct,
	year = {1991},
	pages = {8017--8027}
}

@article{feng_long-range_2025,
	title = {Long-range crossed {Andreev} reflection in a topological insulator nanowire proximitized by a superconductor},
	volume = {21},
	url = {https://www.nature.com/articles/s41567-025-02806-y},
	doi = {10.1038/s41567-025-02806-y},
	number = {5},
	journal = {Nature Physics},
	author = {Feng, Junya and Legg, Henry F. and Bagchi, Mahasweta and Loss, Daniel and Klinovaja, Jelena and Ando, Yoichi},
	month = mar,
	year = {2025},
	pages = {708--715}
}

@article{hofstetter_cooper_2009,
	title = {Cooper pair splitter realized in a two-quantum-dot {Y}-junction},
	volume = {461},
	url = {https://www.nature.com/articles/nature08432},
	doi = {10.1038/nature08432},
	number = {7266},
	journal = {Nature},
	author = {Hofstetter, L. and Csonka, S. and Nyg{\aa}rd, J. and Sch{\"o}nenberger, C.},
	month = oct,
	year = {2009},
	pages = {960--963}
}

@article{schrieffer_relation_1966,
	title = {Relation between the {Anderson} and {Kondo} {Hamiltonians}},
	volume = {149},
	issn = {0031-899X},
	url = {https://link.aps.org/doi/10.1103/PhysRev.149.491},
	doi = {10.1103/PhysRev.149.491},
	number = {2},
	journal = {Phys. Rev.},
	author = {Schrieffer, J. R. and Wolff, P. A.},
	month = sep,
	year = {1966},
	pages = {491--492}
}

@book{cohen1998atom,
	title = {Atom-Photon Interactions: Basic Processes and Applications},
	author = {Cohen-Tannoudji, Claude and Dupont-Roc, Jacques and Grynberg, Gilbert},
	year = {1998},
	publisher = {John Wiley \& Sons}
}

@misc{xue_observation_2025,
	title = {Observation of {Cavity}-{Mediated} {Nonlinear} {Landau} {Fan} and {Modified} {Landau} {Level} {Degeneracy} in {Graphene} {Quantum} {Transport}},
	url = {https://arxiv.org/abs/2506.21409},
	doi = {10.48550/arXiv.2506.21409},
	author = {Xue, Hongxia and Chan, Hsun-Chi and Lin, Zuzhang and Bori{\c{c}}i, Dalin and Zhou, Shaobo and Wang, Yanan and Watanabe, Kenji and Taniguchi, Takashi and Ciuti, Cristiano and Yao, Wang and Ki, Dong-Keun and Zhang, Shuang},
	month = jun,
	year = {2025},
	note = {arXiv:2506.21409}
}

@article{von_delft_superconductivity_2001,
	title = {Superconductivity in ultrasmall metallic grains},
	volume = {10},
	url = {https://onlinelibrary.wiley.com/doi/10.1002/andp.20015130302},
	doi = {10.1002/andp.20015130302},
	number = {3},
	journal = {Ann. Phys. (Leipzig)},
	author = {von Delft, Jan},
	month = mar,
	year = {2001},
	pages = {219--276}
}

@article{garcia-garcia_bcs_2011,
	title = {{BCS} superconductivity in metallic nanograins: {Finite}-size corrections, low-energy excitations, and robustness of shell effects},
	volume = {83},
	url = {https://link.aps.org/doi/10.1103/PhysRevB.83.014510},
	doi = {10.1103/PhysRevB.83.014510},
	number = {1},
	journal = {Phys. Rev. B},
	author = {Garc{\'i}a-Garc{\'i}a, Antonio M. and Urbina, Juan Diego and Yuzbashyan, Emil A. and Richter, Klaus and Altshuler, Boris L.},
	month = jan,
	year = {2011},
	pages = {014510}
}

\end{document}